\documentclass[aps,showpacs,nobibnotes,twocolumn,nobalancelastpage]{revtex4}
\usepackage{graphicx}
\usepackage{amsmath}
\usepackage{latexsym}
\usepackage{dcolumn}
\usepackage{bm}
\input{epsf}
\newcommand{\simgt}{\lower.5ex\hbox{$\; \buildrel > \over \sim \;$}}
\newcommand{\simlt}{\lower.5ex\hbox{$\; \buildrel < \over \sim \;$}}
\newcommand{\be}{\begin{equation}}
\newcommand{\ee}{\end{equation}}
\newcommand{\ba}{\begin{eqnarray}}
\newcommand{\ea}{\end{eqnarray}}
\newcommand{\bi}{\begin{itemize}}
\newcommand{\ei}{\end{itemize}}

\newcommand{\bfi}{\begin{figure}
\epsfxsize=9cm 
\epsffile}
\newcommand{\efi}{\end{figure}}
\newcommand{\bfil}{\begin{figure}}
\newcommand{\efil}{\end{figure}}
\newcommand{\la}{\lesssim}
\newcommand{\mnras}{MNRAS}

\newcommand{\apjs}{ApJS}

\def\mpc{\,h^{-1}{\rm Mpc}}
\def\kmpc{\,h{\rm Mpc}^{-1}}
\def\kpc{\,h^{-1}{\rm kpc}}

\begin{document}
\title{Nonlinearities in modified gravity cosmology I:  signatures of modified
  gravity in the nonlinear matter power spectrum} 
\author{Weiguang Cui$^{1}$}
\email{wgcui@shao.ac.cn}
\author{Pengjie Zhang$^{1}$}
\email{pjzhang@shao.ac.cn}
\author{Xiaohu Yang$^{1}$}
\email{xhyang@shao.ac.cn}

\affiliation{$^{1}$Key Laboratory for Research in Galaxies and Cosmology,
    Shanghai Astronomical Observatory, the Partner Group of MPA, Nandan Road
    80, Shanghai, 200030, China}

\begin{abstract}  A  large  fraction  of  cosmological	 information   on   dark
energy and gravity is encoded in  the  nonlinear  regime.   Precision  cosmology
thus requires precision  modeling  of  nonlinearities  in  general  dark  energy
and modified gravity models.  We modify the Gadget-2 code and run  a  series  of
N-body simulations on modified gravity cosmology to  study  the  nonlinearities.
The modified  gravity  model  that  we	investigate  in  the  present  paper  is
characterized by a single parameter $\zeta$, which  determines	the  enhancement
of particle acceleration with respect to  general  relativity  (GR),  given  the
identical  mass  distribution  ($\zeta=1$   in	 GR).	 The   first   nonlinear
statistics we investigate is  the  nonlinear  matter  power  spectrum  at  $k\la
3h/$Mpc, which is the relevant range for  robust  weak	lensing  power	spectrum
modeling  at  $\ell\la	2000$.	 In  this  study,  we  focus  on  the	relative
difference in the nonlinear  power  spectra  at  corresponding	redshifts  where
different gravity models have the same linear power  spectra.	This  particular
statistics highlights the imprint of modified gravity in  the  nonlinear  regime
and the importance of including the nonlinear regime in testing GR.  By  design,
it is less susceptible to the  sample  variance  and  numerical  artifacts.   We
adopt a mass assignment method based on wavelet to improve  the  power	spectrum
measurement.  We run a series of tests	to  determine  the  suitable  simulation
specifications (particle number, box size and initial redshift).  We find  that,
the nonlinear power spectra can differ	by  $\sim  30\%$  for  $10\%$  deviation
from GR  ($|\zeta-1|=0.1$)  where  the	rms  density  fluctuations  reach  $10$.
This large difference, on  one	hand,  shows  the  richness  of  information  on
gravity in the corresponding scales, and on the other hand,  invalidates  simple
extrapolations	of  some  existing  fitting   formulae	 to   modified	 gravity
cosmology. 
\end{abstract}
\maketitle

\section{Introduction}

One of the biggest challenges of modern cosmology and physics is the existence
of the {\it dark} universe.  Assuming the validity of general relativity (GR),
cosmological  observations lead  to  the  discovery of  dark  matter and  dark
energy, which account for $\sim 96\%$ of the total matter and energy budget of
the  Universe   (e.g.  \cite{Komatsu09}).  However,  since  we   do  not  have
independent tests of GR at relevant scales, the same set of observations could
imply another possibility,  the failure of general relativity  at galactic and
cosmological scales. This possibility, which  serves as an alternative to dark
matter/dark  energy, has  become an  area of  active  research. Discriminating
between the dark matter/dark energy  and modified gravity (MG) models, testing
GR  at cosmological scales  and probing  dark matter  and dark  energy through
cosmological observations,  are thus an entangled task,  of crucial importance
for both cosmology and physics.

Challenges exist in both the observation side and theory side. Although there are
numerous  and  potentially  powerful   observations  suitable  for  this  task
\cite{DETF,  Jain08}, their  precision  measurements are  challenging. On  the
other hand, much  of the cosmological information is  encoded in the nonlinear
regime.  Modeling  the nonlinearities to  the required $\sim 1\%$  accuracy is
challenging  too,  even  for  the  simplest case,  the  standard  $\Lambda$CDM
cosmology with only gravitational interaction (e.g. \citep{Heitmann,H08,Heitmann09,Coyote3}).
Cosmologies  based  on  dynamical  dark   energy  or  MG  are  facing  similar
requirements. References \cite{McDonald06,Linder05,Francis07,Ma99,Heitmann09,Coyote3}
have performed N-body simulations for dynamical and coupled dark energy
models \citep{Baldi}. 

Comparing  to  the  dark   matter/dark  energy  cosmology,  understanding  the
evolution  of the  Universe  in MG models  is often  more
difficult,  due to  the intrinsically  nonlinear feature  of gravity  in these
models or the existence of extra dynamical fields. Despite these difficulties,
the expansion  history of the Universe  and the structure growth  to the first
order have  been robustly understood for many  of the MG models  such as TeVeS
\cite{TeVeS1,TeVeS2}, DGP(short for Dvali,Gabadadze and Porrati) \cite{DGP1,DGP2}
 and   the   $f(R)$   gravity
\cite{fR1,Zhang06,fR2,quasi-static,Nojiri}.   People have  also  achieved success  in
understanding the  nonlinear evolution through  analytical and semianalytical
methods (e.g. \cite{PPF,Koyama09,Scoccimarro09,Cognola1}). Recently, self-consistent
gravity solvers for $f(R)$ \cite{fRsimulation1, fRsimulation2,Hiroaki,Cognola2} and DGP
gravity   \cite{DGPsimulation}  models   have  been   developed  and   led  to
significantly improved understanding of the nonlinear evolution. Simulations
with extra scalar fields and interaction with dark matter 
have also been performed (e.g. \cite{LZ}).  

Since deviations from  GR in general lead to  nonlinear differential equations
of gravity, in principle we have to develop the suitable N-body codes for each
viable MG  model and run  the corresponding simulations.  However,  since we
do not have the final theory of gravity 
based on the first principles, there are in principle infinite MG models to be
investigated.   One possibility  to circumvent  this  problem is  to choose  a
suitable  parameterization  for the  MG  models and  run  a  finite number  of
simulations  to sample  the  relevant parameter  space.  We are  then able  to
interpolate/extrapolate the  simulation results to explore  the whole relevant
parameter space.

The  statistics we focus  on is the  matter power
spectrum. It determines the lensing power spectrum and is also highly relevant
to  the 2D  galaxy clustering,  both  will be  measured to  high precision  by
ongoing and planed imaging surveys,  such as DES, LSST, JDEM/SNAP, Euclid/DUNE,
and KDUST. As we will show later, the density evolution is determined by a
single parameter of gravity, $\zeta$, which quantifies the ability of mass
concentration to distort the space-time metric.   In principle,
$\zeta$ can be both scale, time, and environmental dependence.  Comprehensive
investigation on general $\zeta$ is beyond the scope of the current paper.
Instead, we will adopt a highly simplified form  of $\zeta$  and  run a series
of N-body simulations  to quantify the nonlinear evolution of  the Universe.    

As shown in Heitmann  et al. 2008,  (hereafter H08, \citep{H08}), 
however, to run simulations and model nonlinear matter power spectrum to $1\%$
accuracy to $k\sim 1 h$Mpc is  very challenging, requiring $\rm Gpc$ or larger
simulation box size,  $1024^3$ or more particles, and  beyond.  Being aware of
these difficulties and limited computation resource, we take a modest goal, to
quantify the {\it influence} of MG on the nonlinear matter power spectrum with
respect  to the  standard $\Lambda$CDM  to  $\sim 1\%$  accuracy. Namely,  the
statistics that we will focus on  is the {\it relative difference} between the
nonlinear matter power spectra in the  given MG model and in $\Lambda$CDM.  We
will choose the right redshifts of simulation output such that the linear
matter power spectra in the given MG models are equal to the ones in the
corresponding  $\Lambda$CDM.   This particular  statistics has a number of
attractive features. First, it isolates and highlights the role of MG in
nonlinear evolution. Second, it reduces much of the numerical  artifacts by
taking ratios. Similar tricks have been adopted  in many  previous  simulations
(e.g. \cite{Linder05,McDonald06,Francis07}).   Third, it
can improve the efficiency to understand nonlinearities in MG models, which
can now be reduced to two separate ingredients:  the
nonlinear evolution in $\Lambda$CDM and the relative difference between MG
models $\Lambda$CDM.  
 
The current paper only analyzes a very limited set of
simulations. Nonetheless, it robustly show that, even after scaling out the
difference in the  linear evolution, gravity still leaves significant features
in the nonlinear power spectrum.  In  subsequent  studies,  we  will  run
simulations covering larger parameter space to better understand these
features and hopefully develop a general fitting formula.
Furthermore, we  will study the peculiar
velocity   power   spectrum   and   the   redshift   distortion   (3D   galaxy
clustering), based on these simulations. The ongoing spectroscopic  redshift
surveys like BOSS, LAMOST and 
WiggleZ, and planned spectroscopic  redshift surveys like BigBOSS, JDEM/ADEPT,
Euclid/SPACE and  SKA are  able to measure  these statistics  to unprecedented
accuracy.   We will  also  investigate the  halo  statistics, one  of the  key
scientific goals for galaxy and cluster surveys.

This paper is organized as follows. In Sec. \ref{sec:layout},  we present the MG
parameterization adopted for the simulations, the precision requirements and
the code specifications. In Sec. \ref{sec:tests}, we test the accuracy of the
simulations.  We present major simulation results in Sec. \ref{sec:sr} , 
discussion in Sec. \ref{sec:conclusion} and more results in the Appendix.

\section{The simulation layout}
\label{sec:layout}
\subsection{The $\zeta$ parameterization on modified gravity}\label{sec:qsim}

There  are several existing  parameterizations and general guidances on modified
gravity \cite{MMG, Huterer,  Uzan06, Caldwell07,  Zhang07, Amendola07,  PPF,
  Jain08,Skordis09}.  What we adopt in this paper is the $\zeta$
parameterization. It is a condensed version of
the $G_{\rm eff}$-$\eta$ parametrization \cite{Zhang07,Jain08},
which quantifies two key aspects of gravity.  

 To understand  this point,  we begin with the structure formation in GR, for
 which the central 
 issue is to determine the particle acceleration given the mass distribution. 
The scalar perturbation  of  the  space-time  metric  is  described  by  two  potentials,
$ds^2=-(1+2\psi)dt^2+a^2(1+2\phi)d{\bf  x}^2$.  The  usual  Poisson  equation,
$k^2\phi=4\pi  Ga^2\delta  \rho$ (in  Fourier  space),  relates the  potential
$\phi$ to  the matter distribution  $\rho$.  However, $\phi$ is not the
potential directly responsible for the structure formation in our N-body simulations of
nonrelativistic cold dark matter particles. The contribution to  the particle
acceleration  from  this  potential  is  suppressed  by  a  factor
$(v/c)^2\ll 1$, comparing to  the contribution from the  other potential
$\psi$.  Thus for the  nonrelativistic cold dark matter particles that our
simulations deal with, their  acceleration is determined solely  by $\psi$,
$d(a{\bf  v})/dt=i{\bf k} \psi$,  where ${\bf  v}$ is  the proper  motion. GR  predicts
$\psi=-\phi$, if dark  energy anisotropic stress is negligible.  Now, given an
initial mass distribution $\rho$, we  obtain $\phi$ from the Poisson equation,
with the  coupling constant  $G$. Then through  the relation  $\psi=-\phi$, we
obtain  $\psi$ and  then the  acceleration. Thus  given the  initial positions
(density)  and  velocities  of  particles,  we  can  move  particles  in  each
simulation  time  step  and then  have  a  closed  procedure to  simulate  the
evolution of the Universe under gravity.

A natural parametrization of modified gravity is thus to replace the Newton's 
constant $G$ by the effective Newton's constant $G_{\rm eff}$ and the relation 
$\psi=-\phi$ by $\eta\equiv  -\phi/\psi$. Now, given the mass distribution,
the  acceleration   is  solely  determined  by   the  combination  $\zeta$
\footnote{$\zeta$ defined here is equivalent to $\zeta$ in
  \cite{Jain08},  up to a constant factor of $4\pi G$. },
\be 
\zeta(k,z)\equiv \frac{G_{\rm eff}(k,z)/G}{\eta(k,z)} \ .  
\ee
This is the quantity that enters into the $\psi$-$\rho$ relation,
\be
\label{eqn:psi}
k^2\psi=-\zeta4\pi G \delta\rho\ .  
\ee
GR  has  the value  $\zeta=1$.  Clearly, if  the  two  universes have  the
identical   initial  conditions,  identical   expansion  rate   and  identical
$\zeta(k,z)$, the statistics  of the density and velocity  fields would be
identical \footnote{However, this does not mean that $\zeta$  and  the expansion  rate
uniquely fix the gravitational lensing statistics, since gravitational lensing
is determined by the  combination $\phi-\psi\propto G_{\rm eff}(1+1/\eta) \rho
= \zeta \rho(1+\eta)$. Since this relation is algebraic, there is no extra
simulations required  to evaluate the lensing statistics.  The above arguments
hold as long as $G_{\rm eff}(k,z)$ and $\eta(k,z)$ are deterministic functions
of scale $k$ and redshift $z$.}.  

Thus,  instead of running  a series  of simulations  on a  2D grid  of $G_{\rm
  eff}$-$\eta$ parameter space, we just need to run a series of simulations on
a  1D grid  of $\zeta$  parameter  space. It  significantly reduces  the
amount of  simulations required. This is the major reason that we adopt this single
parameter parametrization on modified gravity. Besides it, there are a number
of attractive features of this parametrization. 

First of all, many MG models, such as  the  DGP  gravity
model \cite{DGP2} and the  $f(R)$  gravity model \cite{Zhang06,fR2} (in  the
linear regime), the Yukawa-like MG model and the $\gamma$-index MG model
\cite{MMG} fit into  this parameterization.  Second,
such  parameterization requires  minimum  modification in  the N-body  gravity
solver, does not require extra computation time, and thus is suitable for fast
exploration of the vast parameter space  of MG models.  In fact, there already
exists a number of  simulations on Yukawa-like gravity \cite{Yukawa,Stabenau}.
Third, $G_{\rm eff}$  and $\eta$ (and hence
$\zeta$), can be  measured in a 
rather   model  independent   manner,   by  combining   imaging  surveys   and
spectroscopic  surveys   \cite{Zhang07,Zhang08}.   This  links   theories  and
observations  directly.   Furthermore,  the  reconstruction  accuracy  can  be
improved  by including  all available  data and  performing  a multiparameter
fitting \cite{Zhao09,Guzik09}. 

Clearly, this  parameterization does not capture  all features of  MG, such as
the  environmental dependence of  gravity, as  found in $f(R)$ gravity
\cite{Chameleon} and the DGP model \cite{Vainshtein}.  Nevertheless, the
simulations  based on  this 
parameterization serve  as an  useful step toward  better understanding  of MG
cosmology. The simulation results can  be used as templates to understand more
complicated  MG  models.  A  close  analogy  is  the  scale-free  simulations.
Although  the   real  CDM(cold dark matter)  transfer  function  is   certainly  not  scale-free
(power-law),  these  scale-free   simulations  do  significantly  improve  our
understanding  of the nonlinear  evolution of  structure formation.  They are
helpful in  developing fitting formula  like that of  Peacock-Dodds (hereafter
PD96, \cite{PD96}) and Smith et al. 2003 (hereafter halofit, \citep{Sm03}). We
hope that similar procedure applies to the case of MG models. For example, the
formalism  proposed by  \cite{PPF} relies  on the  interpolation  between the
nonlinear  power  spectrum  in  GR  and  the one  in  MG  without  environmental
dependence. Understanding the nonlinearities  in MG models without environmental
dependence thus serves as a  natural step to understand nonlinearities in more
complicated MG models.
 
Modifying  existing N-body codes  to incorporate  the $\zeta$ parameterization is
straightforward. The only modification is to change  the particle acceleration
$\vec{a}$ to $\zeta\times \vec{a}$. In the simulation setup,  we fix the expansion
rate identical  to that of  the flat $\Lambda$CDM cosmology \footnote{This is  not an
  arbitrary choice, as it  appears to be. In general  dark energy cosmology
  and  MG cosmology, the  expansion  history and  the  structure formation
  are  independent. Since  the 
structure formation  is affected by quantities  such as the  dark energy sound
speed, the anisotropic stress, $G_{\rm  eff}$ and $\eta$ (and $\zeta$)  at
sub-horizon scales  do not  affect the  expansion. It is  thus natural  to fix
the  expansion rate matching the  observations today  and explore the
possible difference  in the structure growth. }.  In  addition, we do  not aim
to  explore the whole space of $\zeta(k,z)$.  Rather,  we will focus on very
special cases 
of $\zeta$ and postpone the  general investigation for future studies. The
$\zeta(k,z)$   adopted   in   our   simulations   is   scale   independent
($\zeta(k,z)=\zeta(z)$). The success of CMB (cosmic microwave background) and
BBN (big-bang nucleosynthesis) implies that GR is
likely valid in the early Universe.  For this reason, we adopt a step function
in   $z$,    such   that   $\zeta=1$    at   $z\geq   z_{\rm    MG}$   and
$\zeta=$constant$\neq 1$ at $z<z_{\rm MG}$. Throughout this paper, we have
adopted  $z_{\rm  MG}=z_i=  100$,  where  $z_i$ is  the  initial  redshift  of
simulations.  Since  we have GR  valid at high  redshift ($z \geq 100$),  the
transfer  function  at $z_i=100$ adopted in the MG models is identical to that
in GR.  For the adopted MG parameterization,  the linear density growth factor
$D(k,z)$ is scale independent $D(k,z)=D(z)$.  Thus the  linear power  spectrum
for modified gravity models  only   differs  from $\Lambda$CDM  by the  linear
density growth factor  $D(z)$.   In a companion paper, we will explore MG
models with other redshift dependence.

\subsection{The precision requirements}
\label{subsec:precision_requirement}

All MG  simulations begin with  the identical initial condition  at $z_i=100$.
Since the adopted $\zeta$ is  scale independent, the linear density growth
factor $D(z,\zeta)$ is scale independent, as can be seen from the equation
at $z<z_{\rm MG}$,
\be
\delta_m^{''}+\delta_m^{'}\left[\frac{3}{a}+\frac{H^{'}}{H}\right]-\zeta\times
\frac{3}{2}\frac{\Omega_0H_0^2}{H^2a^3}\frac{\delta_m}{a^2}=0 \ .
\label{eqn:lin}
\ee
Here, $^{'}\equiv  d/da$ and $^{''}=d^2/da^2$.  $\Omega_0$ is the  present day
matter density in unit of the  critical density. $H_0$ and $H$ are the present
day Hubble constant  and the Hubble parameter at  $z=1/a-1$. $\delta_m$ is the
linearly  evolved matter over-density  and $D\propto  \delta_m$ is  the linear
density  growth  factor.  Thus,  given   a  redshift  $z_S$  in  the  standard
$\Lambda$CDM, we can find the corresponding redshift $z_{\zeta}$ in the MG
universe, such that
\ba
\label{eqn:z_S}
D(z_S,\zeta=1)&=&D(z_{\zeta},\zeta)\ .
\ea
Here, the  subscript $S$ denotes  the standard $\Lambda$CDM  cosmology.  Since
all  the simulations  begin with  the identical  initial condition,  the above
relation means that,
\ba
P_L(k;z_S,\zeta=1)&=&P_L(k;z_{\zeta},\zeta)\ . \nonumber
\ea
Here $P_L$ is the linear matter  power spectrum. Throughout this paper, we use
the subscript ``L'' for the linear statistics and the subscript ``NL'' for the
nonlinear statistics.

Modifications in GR  change the structure growth history.  The structure grows
faster in  a universe with bigger  $\zeta$.  The primary  quantity that we
want to measure through the simulations is
\be
\label{eqn:epsilon}
\epsilon(k;z_{\zeta},\zeta)\equiv \frac{P_{\rm
    NL}(k;z_{\zeta},\zeta)}
 {P_{\rm NL}(k;z_S,\zeta=1)}\ .
\ee 

$\epsilon\neq 1$ has a number of implications.  (1) If the  nonlinear power
spectrum is  completely determined by  the linear one, independent of the 
 expansion and structure growth history  and the underlying
gravity, then  $\epsilon=1$. A  number of fitting  formulae applicable to GR
have  been extended to  study  the nonlinear  evolution in  MG 
models, based on this assumption. Thus $\epsilon$  provides a direct test on 
the applicability of these
fitting formulae  to MG  models.  Precision cosmology  requires that,  only if
$|\epsilon -  1|\la 10^{-2}$ in the  relevant $k$ range,  may the systematical
error  induced   by  these   fitting  formulae  be   subdominant.  Otherwise,
significant modifications shall be made.  (2) $\epsilon\neq 1$ also means that
there is extra information of gravity encoded in the nonlinear matter power
spectrum, which does not show up in the linear power spectrum at the same
epoch.  This helps to test GR at nonlinear regimes. Such information is
complementary to those in the linear power spectrum at the same epoch and
those in the deeply nonlinear regime where gravity reduces to GR through
environmental dependence mechanisms like the chameleon mechanism and the
Wainshtein mechanism  \cite{Beynon09}. 

Much of  the cosmological  information in weak  lensing surveys come  from the
lensing power  spectrum measurement  at $\ell\la 2000$  of source  galaxies at
$z_s\simeq 1$.   Since the lensing  kernel peaks at half way  between the
source  and   the  observer,  the   peak  contribution  comes   from  $k\simeq
\ell/[\chi(z_s)/2]\la 2  h/$Mpc. At $\ell=2000$, the statistical  error in the
lensing power spectrum measurement can reach below $1\%$ for the planning of wide
surveys.   Under the Limber  approximation, the  lensing angular  (2D) power
spectrum  is   linearly  proportional  to   the  3D  nonlinear   matter  power
spectrum. Thus,  to match  the observation  accuracy, we set  a goal  to model
$\epsilon$ to $\sim 1\%$ accuracy at $z\sim 0.5$ and $k< 3 h/$Mpc.

Since the  simulations run  from the identical  initial condition,  the cosmic
variances in the  resulting power spectra $P_{NL}$ of  different MG models are
highly (positively)  correlated.  Since  the simulations are  run by  the same
code, with the same time steps, errors induced by the numerical artifacts into
$P_{\rm NL}$  should also be  highly (positively) correlated. When  taking the
ratio  of two  power spectra  to evaluate  $\epsilon$, much  of the  errors in
$P_{NL}$ cancels. We thus expect higher accuracy in $\epsilon$ than in $P_{\rm
NL}$. Thus, once  we control the error in $P_{NL}$ to  $\sim 1\%$ accuracy, we
are likely able to measure $\epsilon$ to $1\%$ accuracy.

We run a set of $N=512^3$ particle N-body simulations using the GADGET-2 code,
on the 32-CPU Itanium server at the Shanghai astronomical observatory.  All
the simulations that we use  to calculate $\epsilon$ adopt $L=300 h^{-1}$ Mpc.
Adopting a smaller  box  size  allows  us  to  go  deeper  into  the  nonlinear
regime. However, a smaller box size can  cause numerical artifacts,  due to the
missing of power at $k<2\pi/L$,  which affects the nonlinear evolution through
mode coupling \citep{H08}.   Another reason that  we do not adopt a smaller box
size  is that,  we plan  to use  the same  simulations for  velocity  and halo
statistics, which prefer a larger box size.

\subsection{The GADGET-2 simulation specifications}

We adopt  a parallel GADGET-2 N-body code  \cite{Springel05,Springel01} to run
the simulations.  With a TreePM  algorithm, where only short-range  forces are
computed  with the  ``tree'' method while  long-range forces  are  determined by
particle  mesh (PM)  algorithm, GADGET-2  combines high  efficiency  with high
resolution.

The background  expansion history is fixed  as the one in  a flat $\Lambda$CDM
cosmology with  the matter  density $ \Omega_0  = 0.276$ and  the cosmological
constant $\Omega_{\Lambda}  = 0.724$.  The transfer function  is fixed  by the
above parameters, the baryon density  $\Omega_b = 0.046$ and the dimensionless
Hubble  constant $h =  0.703$. The  amplitude of  the initial  fluctuations is
fixed such that, if linearly evolved  to $z=0$ in the adoption of $\Lambda$CDM
cosmology, the rms  density fluctuation within a sphere of  radius $8 \mpc$ is
$\sigma_8 = 0.811$.

We use $512^3$ PM mesh grids through all the simulations.  The force softening
length $\gamma$  depends on the  mean inter particle separation,  with $\gamma
=0.022L/N^{1/3}$,  where  $L$  is  the  box  size  and  $N$  is  the  particle
number. For  simulations performed with  $512^3$ particles in the $300  \mpc$ box,
$\gamma=12.89 \kpc$.  In GADGET-2, the adaptive  time step is set by $\Delta t
= \sqrt{2\xi \gamma/|a|}$, where $\xi$  controls time step accuracy and $a$ is
the acceleration.   $\xi$ is fixed at  $0.5\%$ for all  the simulations.  With
the adopted  small softening length, the  number of total  adaptive time steps
for our $\Lambda$CDM simulation is about  4000. Fig. 13 of H08 shows that, for
3000 time steps in total,  the resulting difference  in the power  spectra is
less than $0.04\%$. We thus  believe that, the time stepping we adopt suffices
for the purpose of this paper.

\section{Simulation tests}
\label{sec:tests}

In  this section, we  present steps  to control  the robustness  of simulation
results.   We adopt  the  Daubechies  mass assignment  method  to improve  the
accuracy of  power spectrum measurement. We  run a number of  tests to justify
that the  adopted simulation  specifications (particle number,  simulation box
size  and initial  redshift) are  adequate  to constrain  the nonlinear  power
spectrum out to $k=3 \kmpc$ with  $\sim 1 \%$ accuracy.  Finally, we show that
the  modified GADGET-2  code reproduces  the correct  linear evolution  in the
linear regime.

\subsection{Calculating the matter power spectrum}

Usually people  use the fast Fourier  transform (FFT) to  calculate the matter
power spectrum. This requires assigning simulation particles to uniform grids
first.   For  commonly  used  mass  assignment methods,  the  resulting  power
spectrum is biased by the smoothing  and aliasing effects, even at scales well
below  the Nyquist  frequency (e.g.  \citep{Jing05}).  To  reach  the required
accuracy,  we  must  correct  for  these biases. Reference \citep{Jing05}  proposes  an
iterative method  to perform  such task. Alternatively,  \citep{cui2008} adopts
the  Daubechies wavelet  transformation  for the  mass  assignment. The  scale
function of the Daubechies wavelets transform has compact top-hat like support
in  the   Fourier  space,  which   avoids  the  sampling  effect   and  allows
computationally  efficient  mass  assignment  onto grids.   Using  this  scale
function to do the mass assignment  allows for robust measurement of the power
spectrum to $k=0.7 k_{\rm Ny}$  \cite{cui2008}. Throughout this paper, we will
adopt this method to calculate the matter power spectrum.

\subsection{Particle number}\label{sec:check1}

\bfi{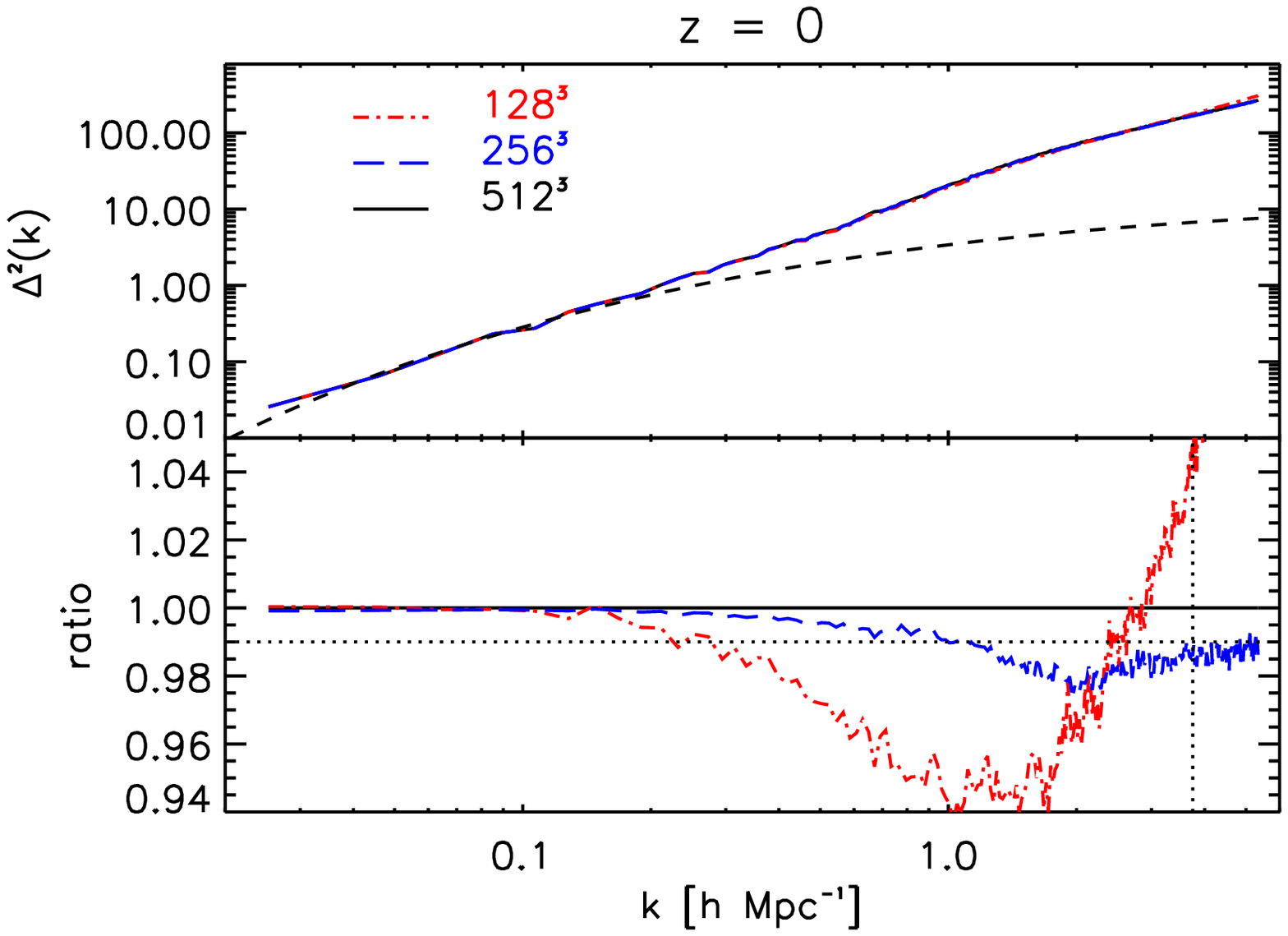}
\caption{The top panel shows the matter power spectra for simulations of different
  particle numbers. The bottom panel  shows the ratios  with respect  to $512^3$
  particle simulation.   All the power  spectra are calculated through  FFT on
  $512^3$ meshes.   The dotted  vertical line shows  the scale of  $0.7 k_{\rm
  Ny}$ for our FFT power spectrum measurement. The black long dashed line is the
  linear power spectrum. At $z=0$, nonlinear correction becomes significant at
  $k> 0.2\kmpc$.} 
\label{fig:mr}
\efi

The particle number in GADGET-2 controls the mass resolution, force resolution
and the time  step. A larger particle number is necessary  to avoid errors from
discreteness  effects at small  scales of  interest. As  pointed out  by Sirko
\cite{Sirko2005}, although  simulations can probe the  evolution of structures
beyond the  particle Nyquist frequency,  $k_{\rm Ny, p}=\pi N^{1/3}/L$,  it is
unclear whether  or not the shot  noise term beyond this  frequency already in
the initial  condition will impact power  at the wavenumbers  of interest. The
issue  may  be  made moot  merely  by  using  negligible  values of  $V/N$  in
simulations.   How many  particles  are required  to  sufficiently sample  the
density field and calculate the matter power spectrum robustly to $k=3 h/$Mpc?
To  answer this  question, we  run  three simulations  with identical  initial
conditions and a box size of $300 \mpc$, but with  $128^3$, $256^3$ and $512^3$
particles, respectively.  Fig. \ref{fig:mr} shows the  nonlinear power spectra
calculated  by the  Daubechies' mass  assignment method.   We see  clearly the
impact of  particle number  on the simulated  power spectrum in  the nonlinear
regime.  The relative difference between the $256^3$ and $128^3$ results at $1
h/$Mpc$\la k\la 3  h/$Mpc is $\sim 4\%$, implying a minimum  error of $4\%$ in
the $128^3$  particle simulation,  due to the  resolution limitation.  But the
relative difference reduces to below $1$-$2\%$ between the $512^3$ and $256^3$
ones,  showing that  the  resolution  induced error  in  the $256^3$  particle
simulation is  reduced significantly. This  trend of convergence  implies that
the resolution induced  error in the $512^3$ simulation  is likely below $\sim
1\%$.  We then  speculate that, if the Daubechies'  mass assignment method was
adopted,  nonlinear power  spectrum  in the  $512^3$  particle simulation  can
attain $O(1\%)$ accuracy  out to $k\sim 3\kmpc$.  To  robustly test it, higher
resolution  simulations  (e.g.  ones  with  $1024^3$ particles  or  more)  are
required.  This test shall be performed in future works.

\bfi{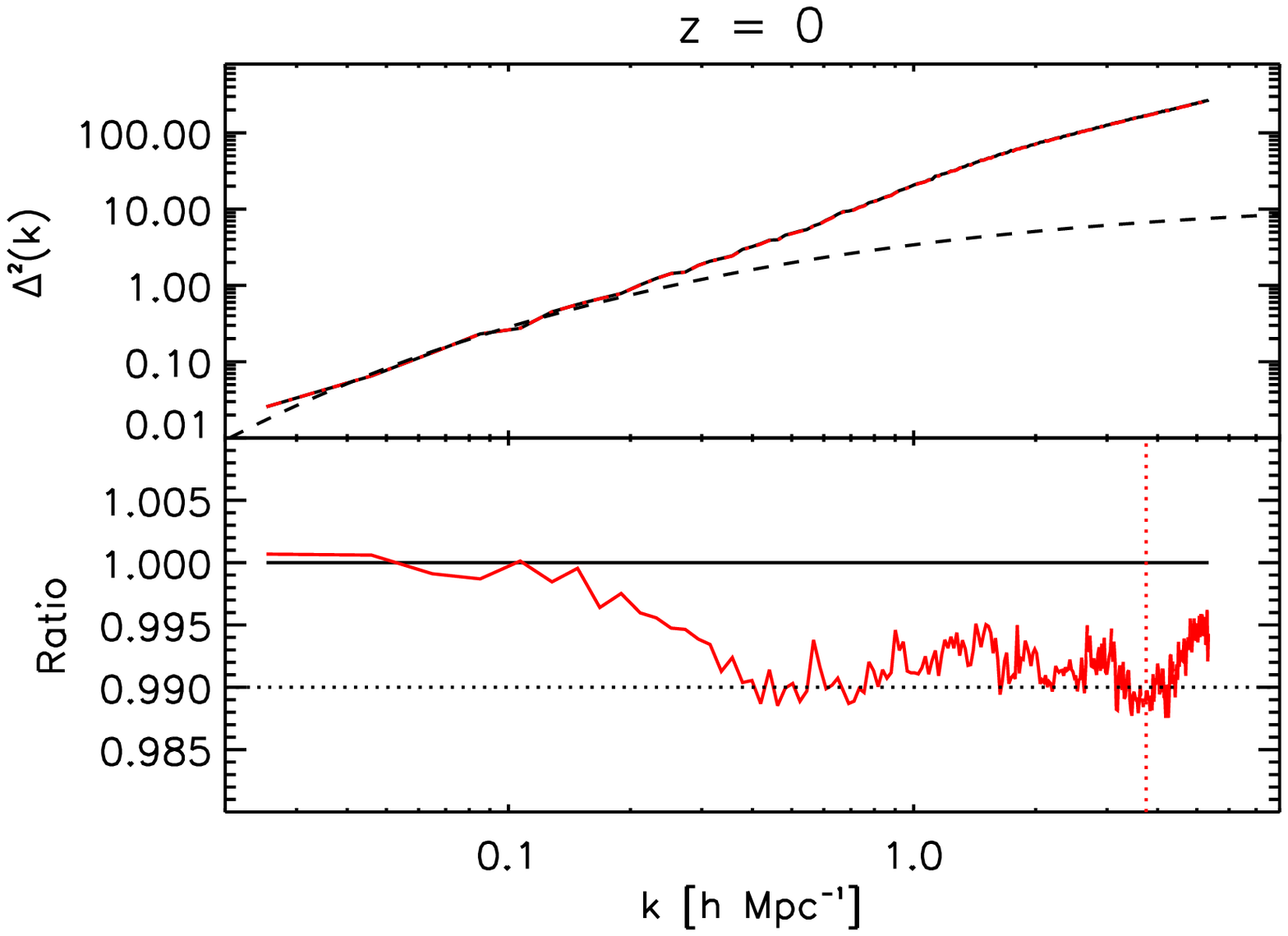}
\caption{The impact of initial redshift  on the matter power spectrum.  In the
  top panel,  we show the power  spectra of the two  simulations with starting
  redshift $z_{i} =  49 $( red line)  and $z_{i} = 100$ (the  black line). The
  bottom panel shows the relative difference. The black horizontal dotted line
  shows the $1\%$ precision requirement.  The red vertical dotted line is $0.7
  k_{\rm Ny}$. The same as Fig. \ref{fig:mr}, the black long dashed line shows the
  linear power spectrum.}
\label{fig:ir0}
\efi

\subsection{Initial redshift}\label{sec:check3}

Testing the  effect of changing the  starting redshift in  simulations is also
important.  Since the  initial  condition is  generated  under the  Zel'dovich
approximation  \citep{Zel}, the  initial redshift  $z_i$ can  not be  too low,
otherwise higher order  corrections can be non-negligible. However,  it is not
automatically the case that higher $z_{i}$ is better, because numerical errors
(most obviously  suppression of power  by limited force resolution)  have more
time  to accumulate  in that  case \citep{McDonald06}.   Our  initial redshift
tests are started at $z_i = 49$ and $z_i= 100$ respectively, both with $512^3$
particles and a $300 \mpc$ box size.  Fig. ~\ref{fig:ir0} compares the two power
spectra at  redshift $z  = 0$.  The  agreement is  better than $1\%$.  We then
justify the choice of $z_i= 100$.

\bfi{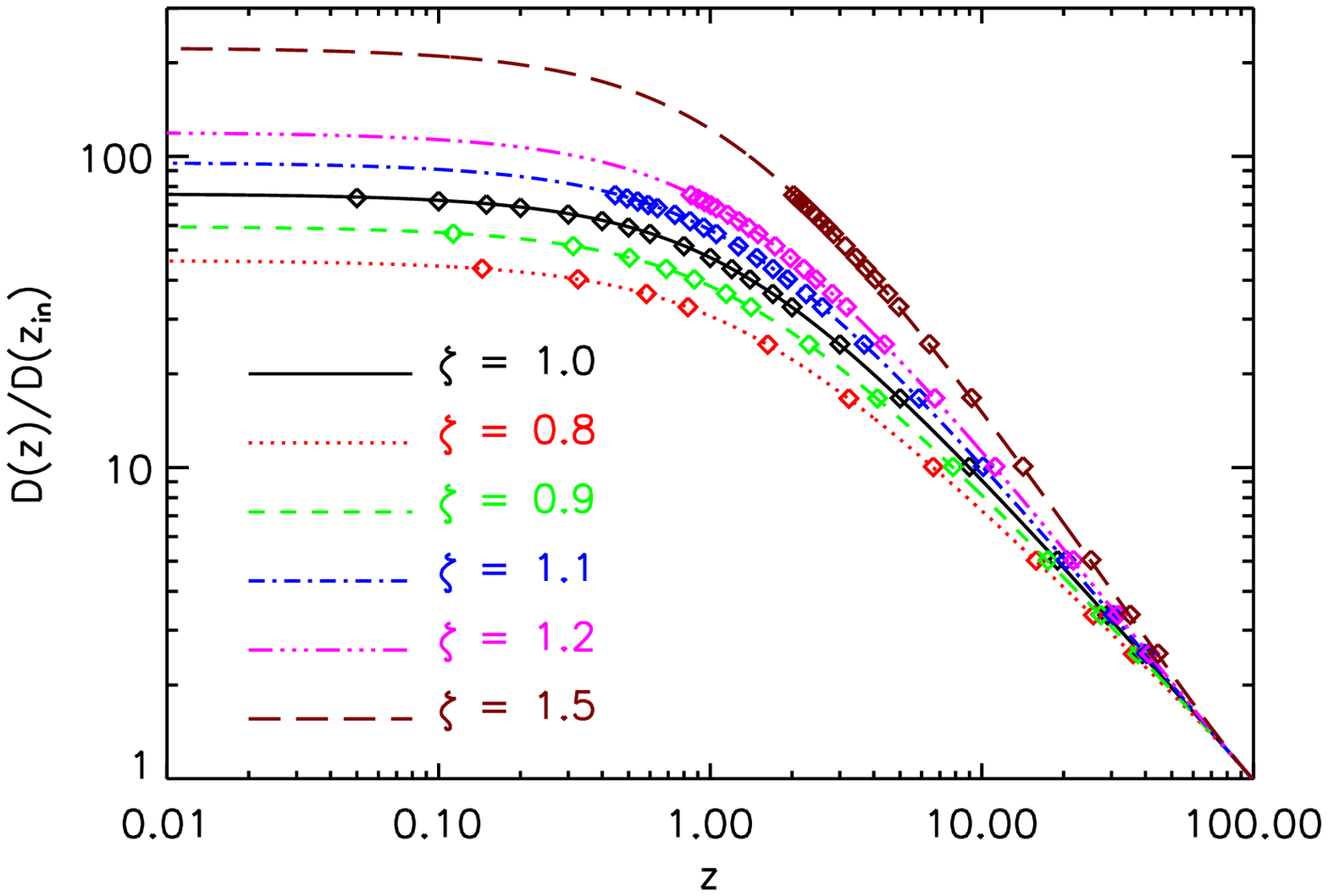}
\caption{The linear  density growth,  normalized at $z=z_i=100$.   The squares
are obtained from the power spectra at $k=0.025 \kmpc$ in the simulations. The
corresponding  redshifts  are shown  in  Table~\ref{t:output}.  The different
colored and style lines are  the theoretical predictions. The agreement  between
the simulated results and theoretical predictions justifies our modified
GADGET-2 code for MG models.}
\label{fig:simcheck}
\efi
\subsection{Simulation box size}\label{sec:check2}

As discussed by  H08 and \citep{McDonald06}, box size  affects nonlinear power
spectrum mainly through two effects.  One is the sample variance.  Smaller box
size simulation suffers larger statistic fluctuations due to fewer independent
modes. Our results are relatively insensitive to this sample variance, because
we take  the ratio of  the power  spectra, which share  more or less  the same
cosmic variance, thus the ratio will cancel much of this sample variance.  The
other  is  systematic  errors  induced  by missing  large-scale  modes,  which
contribute   to  the tidal force. Reference \citep{McDonald06} (see Fig. 9 of
\citep{McDonald06}) shows that the systematic error is substantially small ($<
1\%$ at $z  = 0$) even in box size $L  = 110 \mpc$ out to $k  = 10 \kmpc$.  We
thus think that the $300 \mpc$ box size is suitable. For the adopted box size,
the largest available  mode lies in the linear regime  even at $z=0$, allowing
us to  test the simulation  result against the  linear evolution to  check the
modified GADGET-2 code. 

Based on the above tests, we  justify that, with $512^3$ particles, a $300 \mpc$
box size and initial redshift $z_i = 100$, N-body simulation based on GADGET-2
can help  us reliably  probe the matter  power spectrum to  $k=3\kmpc$.  The
primary statistics that  we investigate in this paper,  $\epsilon$, namely the
ratio of  power spectra of MG models  and $\Lambda$CDM, can reach  unity to an
accuracy of $1\%$. As we will find  later, in MG models,  even for those
with  moderate deviation  from  GR (e.g.   $10\%$  deviation), $\epsilon$  can
deviate  from  unity  to $O(0.1)$,  an  order  of  magnitude larger  than  the
simulation  error. We  thus are  confident  that the  resulting $\epsilon$  is
robust.

\begin{table}[ht]
\begin{center}
\begin{tabular}{|c|cc|c|ccc|}
\hline
$\zeta(z<100)$ &{$0.8$} & {$0.9$} & {$1.0$} &  {$1.1$} &  {$1.2$}
& { $1.5$}  \\
        & &&{\bf $\Lambda$CDM}&&&\\
\hline
&36.08 &  37.60 &  39.00 &   40.27 &  41.49 &  44.52 \\
&25.74 &  27.45 &  29.00 &   30.41 &  31.73 &  35.17 \\
&15.86 &  17.49 &  19.00 &   20.42 &  21.74 &  25.24 \\
&6.623 &  7.830 &  9.000 &   10.11 &  11.20 &  14.18 \\
&3.238 &  4.125 &  5.000 &   5.870 &  6.723 &  9.167 \\
&1.628 &  2.311 &  3.000 &   3.689 &  4.383 &  6.412 \\
&0.828 &  1.414 &  2.000 &   2.586 &  3.179 &  4.949 \\
&0.583 &  1.145 &  1.700 &   2.255 &  2.816 &  4.502 \\
&0.326 &  0.874 &  1.400 &   1.925 &  2.453 &  4.049 \\
&0.145 &  0.689 &  1.200 &   1.703 &  2.210 &  3.745 \\
&      &  0.504 &  1.000 &   1.485 &  1.972 &  3.442 \\
Redshift&      &  0.313 &  0.800 &   1.269 &  1.732 &  3.142 \\
&      &  0.113 &  0.600 &   1.053 &  1.500 &  2.847 \\
&      &0.00719 &  0.500 &   0.947 &  1.383 &  2.700 \\
&      &        &  0.400 &   0.843 &  1.271 &  2.554 \\
& N/A  &        &  0.300 &   0.741 &  1.160 &  2.415 \\
&      &        &  0.200 &   0.639 &  1.053 &  2.281 \\
&      &  N/A   &  0.150 &   0.591 &  1.002 &  2.216 \\
&      &        &  0.100 &   0.541 &  0.951 &  2.152 \\
&      &        &  0.050 &   0.494 &  0.901 &  2.090 \\
&      &        &  0.000 &   0.447 &  0.850 &  2.029 \\
\hline
\end{tabular}
\caption{\label{t:output}The  output  redshifts  for  the simulation  of  each
  $\zeta$.   The baseline  redshifts $z_S$  are  that of  $\zeta =  1$
  ($\Lambda$CDM). The  corresponding redshifts  $z_{\zeta}$ of $\zeta  \neq 1$
  are set up by $D(z_S,\zeta=1) = D(z_{\zeta},\zeta)$.}
\end{center}
\vspace{-0.6cm}
\end{table}

\subsection{Checking Modified GADGET-2}\label{sec:check4}

For the MG  parameterization we adopt Eq. \ref{eqn:psi}, we  only need to do a
minimal  modification to  the GADGET-2  code \citep{Springel05}.  The original
code calculates the gravitational potential  in GR. Multiplying it by a factor
$\zeta$, we obtain the potential $\psi$ in the MG models, which determines
the  acceleration  of nonrelativistic  particles.  In  GADGET-2, because  the
TreePM algorithm  is adopted, the gravitational potential  is explicitly split
into a long-range part and a short-range part. We need to multiply both by the
same factor $\zeta$.  

We test  the modified GADGET-2 code  by comparing the  simulated linear growth
and the theoretically  calculated one. In the linear  regime, the matter power
spectrum $P(k,z)\propto  D^2(z)$. The scale  $k=0.025 \kmpc$ is in  the linear
regime through  all output  redshifts, so we  calculate the power  spectrum at
this  scale   and  compare  it   to  the  theoretical  prediction,   given  by
Eq. ~\ref{eqn:lin}.  Fig. ~\ref{fig:simcheck}  shows the comparisons. The good
agreement indicates that our modified GADGET-2 is correct.

\begin{figure*}
\includegraphics[width=180mm]{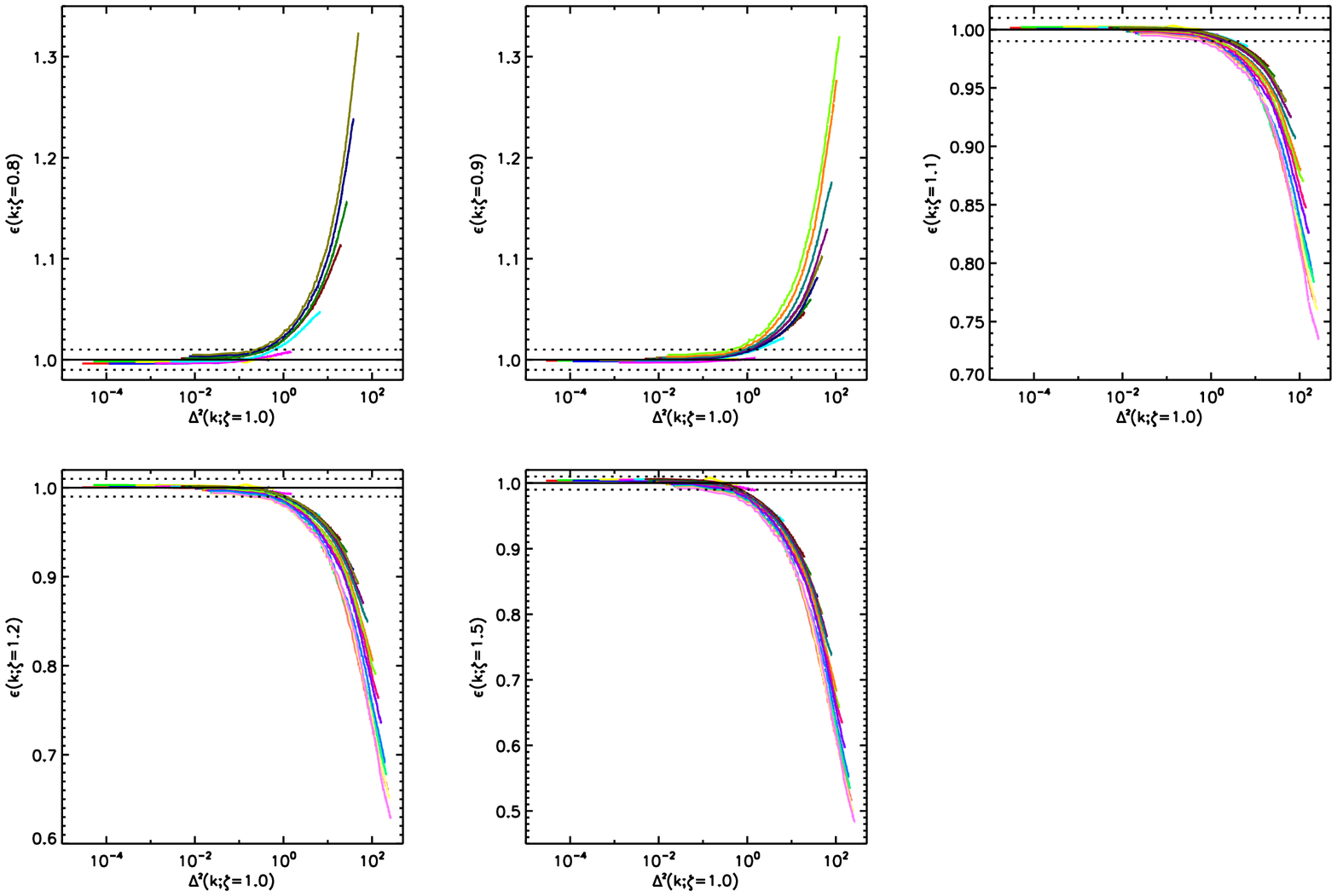}
\caption{ The ratios  of the nonlinear power spectra of MG  models to the ones
in $\Lambda$CDM  when they  have the same linear power.   Results shown in
different  panels correspond to  MG models of  different $\zeta$ values.
Different colors  in each panel show  different output  redshifts. The ones
that reach larger $\Delta^2(k,\zeta=1)$ have lower redshifts.   The  two dotted
black  lines are in the $1\%$ limit. In the bottom-right panel, we show the curves for
$\epsilon$ of different values of $\zeta$ (shown in different line-styles).
This comparison locates at the corresponding redshift with $z_s=1.2$.  [{\it See
the electronic edition of the Journal for a color version of this figure.}] 
}
\label{fig:psall}
\end{figure*}

\section{Simulation results} 
\label{sec:sr}

For the  baseline --- $\Lambda$CDM  simulation, we choose 21  snapshots, whose
redshift  $z_S$  is  shown  in  Table~\ref{t:output}.  We  then  run  five  MG
simulations. We turn on the modified  gravity  at  $z<z_{\rm MG}=100$. This is
certainly not an  unique choice, nor  backed up by a solid argument. For
example, we could turn on the  modified gravity at a much later epoch. This will be the
topic for future study. Naturally,  $\zeta$ adopted in the simulations
should cover the range  allowed by present observations.  Although there is no
direct constraint for this particular type of MG in the literature, current
constraints on MG (e.g. the parameterization investigated by \cite{Mota},
the gravitational slip parameter $\varpi$ in \cite{Daniel} and $\eta$ in
\cite{Bean09} ) imply that, the current constraint on $\zeta$  reaches no better 
than $\sim 10\%$ accuracy.   This instructs us to adopt $\zeta = 0.8, 0.9,
1.1, 1.2$ for the  simulations.  We also run a simulation with
$\zeta=1.5$.  Structure growth in the $\zeta=1.5$ model is very likely too fast to
fit existing observations (Fig. \ref{fig:simcheck}).   Nevertheless, we include this
simulation since dramatic  modifications  in  GR  highlight signatures  of  MG  in  the
nonlinear evolution, lead  to better understanding  of nonlinearity  in MG
models and help  to  improve the  generality  of  fitting formulae based on
these simulations \footnote{We also run  a simulation with $\zeta=0.5$. We
  find that the  structure growth in this model is too linear to be interesting.}.  

The outputs  of these simulations  are chosen according to  Eq. \ref{eqn:lin},
such that the  linear power spectrum of the given  MG model at $z_{\zeta}$
is   identical  to  that   of  $\Lambda$CDM   at  $z_S$.    The  corresponding
$z_{\zeta}$ is  shown in Table~\ref{t:output}. Since  all simulations stop
at $z=0$  and linear density growth  in $\zeta<1$ universe  is slower than
that in  $\Lambda$CDM (Fig. ~\ref{fig:simcheck}),  there will be  no available
$z_{\zeta}$ for comparison in $\zeta<1$ simulations.

The main simulation results are shown  in Fig.  \ref{fig:psall}. As a reminder, the
function $\epsilon-1$  is the relative  difference between the  nonlinear matter
power  spectrum in  the MG  model and  the corresponding  one in  the standard
$\Lambda$CDM [Eq. \ref{eqn:epsilon}].  By design, it scales out the difference
in  the  linear  density  growth   rate  and  thus  highlights  other  factors
determining the nonlinear power spectrum.  Furthermore, due to this particular
design, we are  able to reduce the possible numerical  artifacts and model the
nonlinearity to $1\%$  accuracy.  To better show the  effect of nonlinearity,
we  plot  $\epsilon$ against  the  nonlinear  matter  power spectrum  variance
$\Delta^2(k;z_S,\zeta=1)$. To  illustrate the evolution  of $\epsilon$, we
plot all $\epsilon$ of the same $\zeta$ in the same plot. At the bottom-right panel
of Fig. \ref{fig:psall}, we show a comparison between different values of $\zeta$.
The redshifts for different $\zeta$ of this comparison are selected when they
have the same linear power spectrum as $\Lambda CDM$ simulation at $z_s=1.20$.

\subsection{Signatures of modified gravity in the nonlinear regime}

The results in Fig. \ref{fig:psall} shows significant imprints ($\epsilon \neq 1$) of modified
gravity in the  nonlinear evolution.  Deviation of $\epsilon$ from unity
becomes stronger at smaller scales and lower redshifts where nonlinearity is
stronger. We find that, by proper scaling of $\Delta^2$, 
curves of $\Delta^2$-$\epsilon$ of fixed $\zeta$ can fall upon each other. We
will discuss this   behavior further in the Appendix and
show its application to  develop a fitting formula of $\epsilon$, accurate to
$\sim 1\%$.  Whether or not this behavior
is generic will be investigated in future studies.   

There is another interesting behavior in  the  nonlinear evolution.  The
density  growth in  $\zeta>1$  cosmology is 
faster  (Fig.  \ref{fig:simcheck}).   However,  after scaling  out the  linear
growth,  the  normalized  nonlinear  evolution  is  actually  slower  (namely,
$\epsilon<1$ when $\zeta>1$,  Fig.  \ref{fig:psall}).  When $\zeta<1$,
the  behavior is  opposite ($\epsilon>1$).   

The halo  model may  explain such behavior.  The nonlinear power spectrum can
be decomposed into two terms \cite{halomodel}:
\ba \Delta^2_{\rm NL}(k,z)&=&\Delta^2_L(k,z)\left[\int_0^{\infty}
  M\delta(k|M,z)b(M)\frac{dn}{dM}dM\right]^2 \nonumber\\
&&+\frac{k^3}{2\pi^2}\int_0^{\infty} M^2\delta^2(k|M,z) \frac{dn}{dM}dM \ .
\ea
Here, $\delta(k|M,z)$  is the  Fourier transform of  the density profile  of a
halo  with mass  $M$  at redshift  $z$,  normalized so that  $\delta(k\rightarrow
0)\rightarrow 1$.   The halo abundance is  $dn/dM$, and for  convenience it is
normalized such  that $\int Mdn=1$.  The first  term on the right  hand of the
equation  is the  two-halo term,  which dominates  in the  linear  regime. The
second term is  the one-halo term, which dominates  in the strongly nonlinear
regime.  Since  $\delta(k\rightarrow 0)\rightarrow 1$ is independent  of the halo
density profile and $\int M b(M)dn=1$, the two-halo term is not very sensitive
to  the halo  density profile.   Since $D(z_{\zeta},  \zeta)  = D(z_S,
\zeta=1)$, as  required, the  two-halo term in  the MG model  is (roughly)
equal to the  two-halo term in $\Lambda$CDM, where they  are dominant.  On the
other hand, in the nonlinear regime, the one-halo term strongly depends on the
halo density  profile, which depends  on the structure growth  history.  Since
structure    grows    faster    in    $\zeta>1$   cosmology,    we    have
$z_{\zeta}>z_S$.  Halos in this universe  form in a background with higher
mean  density ($\propto  (1+z)^{-3}$).  We  then expect  them to have a smaller
concentration \cite{DHZhao}.   For the same mass, a smaller concentration means
a smaller $\delta(k|M,z)$  and a smaller contribution from the  one-halo term.  We
then  expect $\Delta^2_{\rm NL}  (k; z_{\zeta},  \zeta) <\Delta^2_{\rm
NL} (k;z_S, \zeta=1)$ and thus $\epsilon(\zeta>1)<1$ in the nonlinear  regime.  
We defer this investigation to a  forthcoming paper, where we will measure and
compare  the  halo   mass  functions  and  profiles  between   MG  models  and
$\Lambda$CDM. For the same  reason, we expect $\epsilon(\zeta<1)>1$ in the
nonlinear regime.  Whether or not the halo model will lead to a satisfying
description of $\epsilon$ and thus the nonlinear power spectrum in MG models
is an interesting project for further investigation. 

In a word, large deviation of $\epsilon$ from unity implies that there is
valuable information of gravity encoded in the nonlinear regime, which is
complementary to those encoded in the linear matter power spectrum at the same
epoch. It will be interesting to quantify how significantly this imprint of
gravity in the nonlinear regime can improve cosmological tests of gravity,

\subsection{Implications on the applicability of some existing fitting formulae}
The particular definition of $\epsilon$ allows us to address a key question in
understanding the nonlinearity, {\it  is the nonlinear power spectrum uniquely
determined by the  linear one at the same epoch?}  Equivalently, if the linear
matter power spectra of two  cosmologies are identical, will the corresponding
nonlinear matter power spectra be identical?

The  influential  HKLM (Hamilton, Kumar, Lu and Matthews) procedure  \cite{HKLM} 
 assumes  so.  It postulates  the
existence of an one-to-one mapping  between the linear correlation function at
a linear  scale and the  nonlinear correlation function at  the corresponding
nonlinear scale. Reference \cite{Jain95} found that this mapping depends on the slope of
the  linear  power  spectrum.  Hence  after, the  slope  dependence  has  been
explicitly  incorporated in  several fitting  formulae, including  the popular
PD96 fitting  formula \cite{PD96}  and the Smith  et al. 2003  halofit formula
\citep{Sm03}. In these  fitting formulae, the mapping is  expressed in Fourier
space,     of     the     functional    form     $\Delta^2_{\rm     NL}(k_{\rm
NL},z)=\aleph(\Delta^2_L(k,z))$.  This mapping is  nonlocal. For  example, in
PD96,    $\Delta^2_{\rm    NL}(k_{\rm   NL},z)$    depends    not   only    on
$\Delta^2_L(k_L,z)$ at the  corresponding linear scale $k_L$, but  also on the
effective power index  $n_{\rm eff}$ at some linear scale,  often chosen to be
$k_L$  or   $k_L/2$.  Furthermore,  the  mapping  has   extra  dependences  on
cosmology.         In          PD96,         $\Delta^2_{\rm         NL}(k_{\rm
NL},z)=\aleph(\Delta^2_L(k,z),g(z))$.  The cosmological dependence  $g(z)$ has
clear  physical   meaning,  $g(z)\propto  D(z)(1+z)$  and   is  normalized  to
$g(z\rightarrow  \infty)=1$.    In  the  halofit,   $\Delta^2_{\rm  NL}(k_{\rm
NL},z)=\aleph(\Delta^2_L(k,z),\Omega_m)$,  where $\Omega_m$  explicitly enters
several fitting parameters.  For a comprehensive review, refer to \cite{Sm03}.

The HKLM procedure and its variations are successful in capturing the
nonlinearities in CDM plus GR simulations. For this reason, they are often
extended to predict the nonlinear matter power spectrum in MG/dark energy
models \cite{Knox05,Ishak05,Wang07,Dore07,Bean09}.  The applicability of the
resulting fitting formulae to MG 
models has been tested against several MG simulations
\cite{Laszlo08,Khoury,Stabenau}.  In general, there is reasonable agreement at
the $\sim 10\%$ level. But discrepancies are also noticed
(e.g. \cite{Stabenau,Khoury}). Our simulations, with improved simulation
accuracy, improved power spectrum measurement method and specifically designed
statistics, are able to identify the discrepancies at the $1\%$ level.

Our  simulation  results   (Fig.   \ref{fig:psall})  show  unambiguously  that
$\epsilon\neq  1$ in  the  nonlinear  regime.  For a $10\%$  deviation from  GR
($\zeta=1.1$, or  $\zeta=0.9$), the resulting  nonlinear power spectra
can differ by  $20\%$-$30\%$ at $\delta \sim 10$  ($\Delta^2\sim 100$), to the
corresponding one  in the $\Lambda$CDM.   The deviation becomes larger  if the
deviation of $\zeta$ from unity  is larger. Quite obviously, the nonlinear
matter power  spectrum is not completely  determined by the linear  one at the
same epoch. Reference \cite{Ma07} demonstrated by the  case of dynamical dark energy,
that the structure growth history is also 
responsible for shaping the nonlinear matter power spectrum. The MG models we
investigate have a different structure growth history (e.g. different structure
growth rate), and this may explain the observed significant deviation of
$\epsilon$ from unity.

Our simulation set up allows us to evaluate the applicability of using several
existing  fitting  formulae  to   MG  models  even  without  directly  testing
them. Since  all the simulations have  identical linear power  spectra and the
present  day  matter  density  $\Omega_m$,  the  halofit  would  then  predict
$\epsilon=1$.   Our simulation result  of $\epsilon$  then implies  that a $\sim
20\%$  error may occur  if one  uses the  halofit to  calculate $\Delta^2_{\rm
NL}(\zeta=1.1)$  (and  $\Delta^2_{\rm  NL}(\zeta=0.9)$)  at  the  over
density $\delta \sim 10$. The application of PD96 to MG models is a little bit
tricky. In PD96, $g(z)$ is the ratio of the linear density growth rate between the
given CDM cosmology and the  $\Omega_m=1$ flat universe. However, this form of
$g(z)$, despite  its clear  physical meaning, does  not apply to  more general
cases,  such  as  the  case  of  dynamical  dark  energy  models  \cite{Ma99}.
Furthermore,  PD96  is based  on  the  stable  clustering hypothesis.   N-body
simulations  show  that  this  hypothesis is  problematic  \cite{Sm03,Jing01}.
Simple extrapolation of PD96 to the MG models should be avoided too.

\section{Discussion and Conclusion}
\label{sec:conclusion}

In this paper, we modify the GADGET-2  TreePM code to run a set of simulations
for parameterized modified  gravity models. As the first paper  in a series, we
focus  on the  nonlinear power  spectrum  in MG  models.  We  take
several steps  to improve/test the  model and simulation accuracy.   First, we
adopt   an  advanced   analysis   method  to   improve   the  power   spectrum
measurement. We  then test  the impact of  various mass and  force resolution,
time step, initial redshift, and box  size on the nonlinear power spectrum, and
find suitable  simulation specifications which meet  our accuracy requirement.
Finally, we focus on a  particular quantity $\epsilon$, the ratio between
the nonlinear power  spectra between MG models and  $\Lambda$CDM with the same
power  spectra  at a large linear  scale.   This quantity  can be measured to higher
accuracy than the nonlinear power spectrum itself, since much  of  the sample
variance  and simulation artifacts  are reduced in $\epsilon$.  By
construction, deviation of $\epsilon$  from unity is a signature of MG
imprinted in the nonlinear evolution. It also  means that the 
nonlinear matter power spectrum is not uniquely fixed by  the  linear one  at 
the  same redshift.  It thus  also represents  the minimum
systematical  error  induced by  simply  extrapolating  some existing  fitting
formulae to these  MG models.  We find that, the  deviation of $\epsilon$ from
unity can reach  $O(0.1)$ where the rms density  fluctuation reaches $10$, for
MG models  with a $10\%$  deviation from  GR.  As an exercise toward a general
fitting formula of this signature of MG, we develop  a simple fitting  formula
of $\epsilon$, accurate to $\sim 1\%$,  working for the particular MG models
that we investigate.

Significant  improvements are  required to  reach precision  modeling  of the
nonlinear matter power spectrum in more  general MG models. In the next steps,
we will run more simulations with larger box sizes, $1024^3$ or more
particles, various initial conditions, and various expansion histories. 

More importantly, we need to explore larger MG parameter space.  For example,
we may need to vary $z_{\rm MG}$ to see its influence. Furthermore, instead of taking
$\zeta$ as a step function with no scale dependence, we shall explore more
complicated time dependent and scale dependent $\zeta$ models. In a companion
paper, we will 
explore the  minimalist MG  model ($\gamma$-index, \cite{Huterer}),  which has
been  adopted  by   the  Figure  of  Merit  Science   Working  Group  (FoMSWG)
\citep{Albrecht}  for forecasting. In  this model,  the linear  density growth
rate is given by $f(a)\equiv d\ln D(a)/d\ln a=\Omega^{\gamma}_m(a)$. 
Here, the  growth index $\gamma$ is a  constant, whose value is  $\simeq 0.55$ in
$\Lambda$CDM. Stage-IV dark energy surveys can
constrain this parameter with a rms error 
$O(0.01)$ (e.g.  \cite{Stril09}).  $\Omega_m(a)=\Omega_0a^{-3}/(H^2/H_0^2)$ is
the matter  density at  redshift $z=1/a-1$. One  particular advantage  of this
parameterization is that,  since $\Omega_m(a\rightarrow 0)\rightarrow 1$, even
for a model with time-constant $\gamma$ can approach GR at high redshift.  

The $\zeta$ parameterization can incorporate this model. The corresponding
$\zeta$ can  be    obtained    from    Eq.   \ref{eqn:lin},    
\be    \zeta=\frac{2}{3}
\frac{f^2+f(2+H^{'}a/H)+af^{'}}{\Omega_m(a)}\ ,\ \  ^{'}\equiv\frac{d}{da}\ .  
\ee  
It is interesting to see
whether new features will arise in  the nonlinear regime and how to extend the
proposed fitting scheme for this MG model.

Its is much harder to simulate  realistic MG models such as $f(R)$ and DGP,
which have complicated environmental dependences.  We hope that, studies on MG
models  without  environmental  dependence  can provide  useful  templates  to
understand these MG models.

\acknowledgements
We thank Volker Springel for the N-genic, public code GADGET-2 and detailed help.
We thank Youcai  Zhang, Raul Angulo, Klaus Dolag and  Till Sawala for valuable
discussions. This work is supported in part by the one-hundred talents program
of the  Chinese academy of science,  the national science  foundation of China
(grant  No.   10533030,  10821302,   10925314  \&  10973027),  the  CAS  grant
KJCX3-SYW-N2 and the 973 program grant No.  2007CB815401 \& 2007CB815402.

\appendix
\section{The fitting formula}
\label{sec:fitting}
\begin {figure*}
\includegraphics[width=180mm]{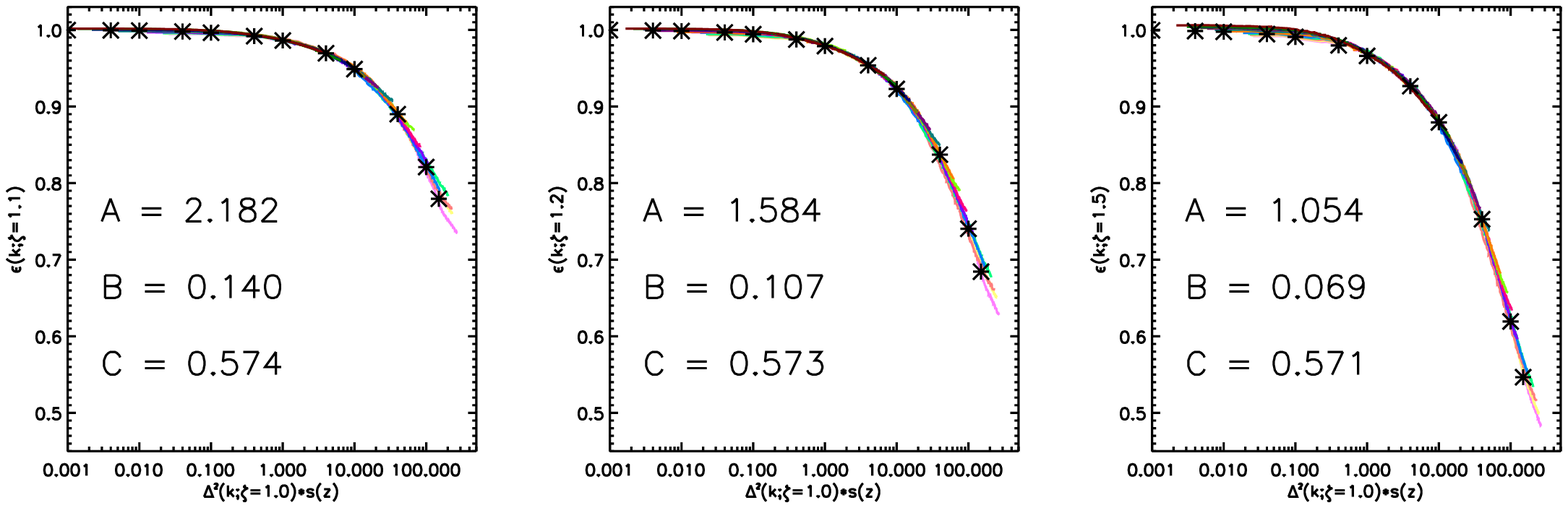}
\caption{Similar  to Fig.  \ref{fig:psall},  but only  shown for  results with
  $\zeta>1$.  In addition, lines  corresponding to different redshifts are
  shifted by a factor of $s(z)$  as described in Eq.  \ref{eqn:s}. These lines
  are then fitted  with three parameters $A$, $B$ and  $C$.  The asterisks are
  the fitting points produced by the fitting recipe.}
\label{fig:allfit}
\end{figure*}

We  demonstrate the feasibility to develop an accurate fitting
formula for  $\epsilon$, which 
quantifies the difference of the  nonlinear evolution between the MG cosmology
and  $\Lambda$CDM.   In combination  with  the  existing $\Lambda$CDM  fitting
formulae, the nonlinear power spectrum in  MG models can be predicted. We only
use  the   results  of  $\zeta>1$  simulations  to   develop  the  fitting
formula. We reserve  the $\zeta<1$ simulations to check  the generality of
our fitting formula. This fitting formula is by no mean applicable to general
MG models. Nonetheless, we hope that it serves as an useful exercise toward 
more general fitting formula, which we will explore elsewhere.

\subsection{Developing the fitting formula}

As  shown in  Fig. \ref{fig:psall},  $\epsilon$ is  a function  of  the scale,
redshift and $\zeta$,
\ba
\epsilon(k,z_{\zeta},\zeta)=u(x,z_S,\zeta)\nonumber \\
\ x\equiv \Delta^2_{\rm NL}(k,z_S,\zeta=1)\ .
\ea
We will develop the fitting formula according to the following steps.

\textbf{Step  1}:   By  visually  inspecting Fig.   \ref{fig:psall}  for  each
$\zeta$, it looks feasible to  move each lines horizontally such that they
fall    upon   each    other.     This   horizontal    shift   indeed    works
(Fig. \ref{fig:allfit}).  Mathematically, this means  that we can find a shift
function $s(z_S,\zeta)$, such that
\be
\label{eqn:f}
u(x,z_S,\zeta)=u(y,\zeta);\ y=x\times s(z_S,\zeta) \ .
\ee
In  the exercise,  we  fix $\zeta$  first.  Then for  each  $z_S$ (or  the
corresponding  $z_{\zeta}$),  we find  the  suitable  $s$,  such that  the
corresponding curve,  after the horizontal  shift, overlaps with the  one with
$z_S=0$. To do so, we have required that $s(z_S=0,\zeta)=1$. We find that
higher redshift  curves should be  shifted more leftward. This  requires that
(i) $s(z_S)$ is positive, and  (ii) $s(z_S)$ monotone decreases with respect to
$z_S$. These requirements  help us to find the  following fitting function for
$s$:
\ba
\label{eqn:s}
s(z_S,\zeta) =
\left[\frac{D(z_S,\zeta=1)}{D(z_S=0,\zeta=1)}\right]^{A(\zeta)}\ ,
\ea
where $A(\zeta)>0$.

\textbf{Step  2}:  The  next step  is  to  figure  out  a suitable  form  for
$u(y,\zeta)$, namely, to fit  those curves in Fig. \ref{fig:allfit}. There
are several guidelines.  (1) $u>0$, since both power spectra must be positive.
(2)  $u(y,\zeta=1)=1$,   by  the  definition.    (3)  $\epsilon=u<1$  when
$\zeta>1$  and $\epsilon>1$  when  $\zeta<1$ (Fig.   \ref{fig:psall}).
These behaviors motivate us to propose the following fitting function:
\be
\label{eqn:u}
u(y,\zeta)=e^{(1-\zeta)B(\zeta)y^{C(\zeta)}}\ .
\ee
As long as $B(\zeta)>0$, all three conditions are satisfied. 

\textbf{Step 3}: For each $\zeta$, we fit the simulation data and find the
best fit $A$, $B$ and $C$, whose values are shown in Fig. \ref{fig:allfit}. We
then need  to find the  suitable form to  model the $\zeta$  dependence of
these  parameters.  The following  functions  with  the associated  parameters
provide a good fit:
\ba
\label{eqn:abc}
A(\zeta)&=& e^{(a_0-\zeta)}\ ,\  a_0 =  1.745\ ,\nonumber\\
B(\zeta)&=& b_0+b_1\zeta^{-4} ,\  b_{0,1}=0.0429,0.133\ ,\\
C(\zeta)&=& 0.573\ .\ \nonumber
\ea

\subsection{Calculating the nonlinear power spectrum in MG models}

We summarize  the procedure to  calculate the nonlinear matter  power spectrum
$\Delta^2_{\rm NL}(k,z_{\zeta},\zeta)$ using our fitting formula.
\bi
\item  For  the given  redshift  $z_{\zeta}$ in  the  MG  model, find  the
  corresponding  $z_S$ in  $\Lambda$CDM through  Eq.  \ref{eqn:z_S}.   The two
  corresponding power spectra at large linear scale are then identical.
\item Calculate  $\Delta^2_{\rm NL}(k,z_S,\zeta=1)$.  This can  be done by
  using either  direct $\Lambda$CDM simulations  (as in our case)  or existing
  fitting formulae such as the halofit.
\item Combining  Eqs. \ref{eqn:f}, \ref{eqn:s},  \ref{eqn:u} \& \ref{eqn:abc},
  we  are  then  able  to predict  $\epsilon(k,z_{\zeta},\zeta)$.   In
  combination with $\Delta^2_{\rm NL}(k,z_S,\zeta=1)$, we can then predict
  the  nonlinear matter  power spectrum  $\Delta^2_{\rm  NL} (k,z_{\zeta},
  \zeta)$ in the given MG model.  
\ei
%

\bfi{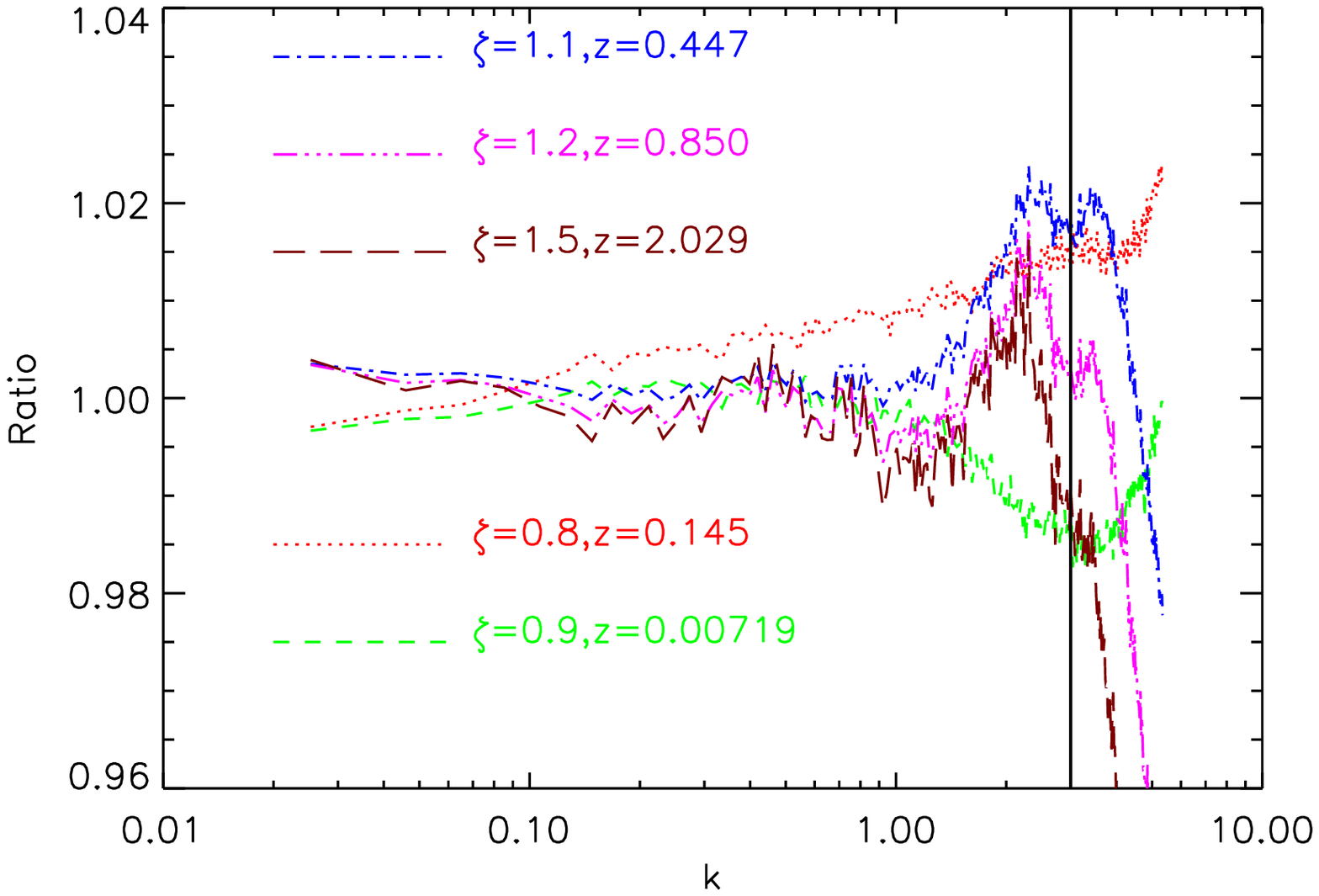}
\caption{Testing  the accuracy  of the  fitting  formula.  We  plot the  power
  spectrum ratios as a function of  $k$ between the simulation results and our
  model  predictions.  Different  $\zeta$ with  its checking  redshift are
  shown in  different colors and  line styles.  The  vertical line is  at $k=3
  \kmpc$, below which our simulation results are reliable, as shown in Sec.
  \ref{sec:tests}. }
\label{fig:expcheck}
\efi
\subsection{Testing the fitting formula}

Becuase of the very limited  simulations that  we have,  we are  not able  to perform
comprehensive tests  against the generality of the  fitting formulae. However,
we are indeed able to check it against our $\zeta<1$ simulations. Since we
do  not  use   these  simulations  to  find  the   fitting  parameters,  these
$\zeta<1$ simulations  can provide an independent check against  our fitting
formula.

In Fig.  \ref{fig:expcheck},  we show the performance of  our fitting formula.
The ratios between the simulated  and predicted nonlinear power spectrum of MG
models are  shown using  different line styles  for different  $\zeta$ and
redshifts as indicated.  In general, our fitting formula is accurate to 1-2$\%$
in the  range $k<3h/$Mpc.  Although  results for the cases  with $\zeta<1$
are extrapolations of  our fitting formula, their performances  are as good as
for  the   $\zeta>1$  cases.   Such  good   performance  demonstrates  the
applicability of our fitting formula.  And  we do not try to seek for possible
slightly more accurate but much more complicated fitting formulae.

We have shown that the  proposed fitting formula provides a good description of
the  nonlinear matter  power spectrum,  for the  specific form  of MG  that we
adopt. Is it applicable to  other cases? We are not able to answer
this question by the existing  simulations.  However, we still want to discuss
a  less general  question, is  it  applicable to  the MG  models with  $z_{\rm
MG}\neq 100$?  In the fitting  formula, the only quantity dependent of $z_{\rm
MG}$  is   $D(z_S,\zeta=1)=D(z_{\zeta},\zeta)$.   This  dependence
alone may not be  sufficient. It is very likely that $A$,  $B$, and $C$ depend
on  $z_{\rm  MG}$,  too.   This  is  a key  issue  for  future  investigation.
Nevertheless, we show that it is possible to develop a fitting formula for MG
models  by the  above simple  technique.  It may  also be  applicable to  more
general  MG   models.  This  is   again  an  interesting  issue   for  further
investigation.

\end{document}